\documentclass[aps,pra,twocolumn,superscriptaddress,showpacs]{revtex4}
\usepackage{amsmath,amssymb,graphicx,url}
\usepackage[hypertex,colorlinks=true,citecolor=blue,urlcolor=blue]{hyperref}

\newcommand{\PR}[4]{
Phys. Rev. #1 {\bf #2}, \href{http://dx.doi.org/10.1103/PhysRev#1.#2.#3}{#3} (#4)}
\newcommand{\PRRC}[4]{
Phys. Rev. #1 {\bf #2}, \href{http://dx.doi.org/10.1103/PhysRev#1.#2.#3}{#3(R)} (#4)}
\newcommand{\PRL}[3]{
Phys. Rev. Lett. {\bf #1}, \href{http://dx.doi.org/10.1103/PhysRevLett.#1.#2}{#2} (#3)}
\newcommand{\RMP}[3]{
Rev. Mod. Phys. {\bf #1}, \href{http://dx.doi.org/10.1103/RevModPhys.#1.#2}{#2} (#3)}
\newcommand{\Nature}[4]{
Nature (London) {\bf #2}, \href{http://dx.doi.org/10.1038/#1}{#3} (#4)}
\newcommand{\Nat}[5]{
Nat. #2 {\bf #3}, \href{http://dx.doi.org/10.1038/#1}{#4} (#5)}

\begin{document}
\title{Magnetization and phase transition induced by \\
circularly polarized laser in quantum magnets}
\author{Shintaro Takayoshi}
\affiliation{National Institute for Materials Science, 
Tsukuba, Ibaraki 305-0044, Japan}
\author{Hideo Aoki}
\affiliation{Department of Physics, The University of Tokyo, 
Hongo, Tokyo 113-0033, Japan}
\author{Takashi Oka}
\affiliation{Department of Applied Physics, The University of Tokyo, 
Hongo, Tokyo 113-8656, Japan}

\date{\today}

\begin{abstract}
We theoretically predict a nonequilibrium
phase transition in quantum spin systems induced by a laser,
which provides a purely quantum-mechanical way of 
coherently controlling magnetization.
Namely, when a circularly polarized laser is applied
to a spin system, the magnetic component of a laser
is shown to induce a magnetization normal to the plane of
polarization, leading to an ultrafast phase transition.
We first demonstrate this phenomenon numerically
for an $S=1$ antiferromagnetic Heisenberg spin chain,
where a new state emerges with magnetization perpendicular to the 
polarization plane of the laser in place of 
the topologically ordered Haldane state. 
We then elucidate its physical mechanism
by mapping the system to an effective static model. 
The theory also indicates that the phenomenon should 
occur in general quantum spin systems with a magnetic anisotropy. 
The required laser frequency is in the terahertz range, 
with the required intensity being 
within a prospective experimental feasibility.
\end{abstract}

\pacs{42.55.Ah, 75.10.Jm, 75.10.Pq, 75.30.Gw}

\maketitle

\section{Introduction}
There is a growing fascination with physics of nonequilibrium systems, 
which is becoming an important topic in condensed-matter 
and other fields of physics. For electron systems, a host of 
novel phenomena induced in nonequilibrium situations, 
such as photoinduced Mott transitions~\cite{Iwai03,Okamoto11,Aoki14}, 
photoinduced topological transitions~\cite{Oka09,Wang13,Kitagawa11,Lindner11}, 
etc, have been fathomed. 
Non-equilibrium physics is explored in cold atoms as well, 
where quantum simulation
is being realized~\cite{Bloch08,Simon11}. 
Now, we pose a question: can we propose novel non-equilibrium 
phenomena for quantum {\it spin systems} 
as opposed to electron systems? 
Spin systems are a distinct class of many-body systems, 
having various possibilities as in recent spintronics, so 
nonequilibrium phenomena are highly intriguing.  
Here we propose a novel, nonequilibrium way 
to {\it coherently} control many-body states in spin systems with intense 
laser fields. 
In contrast to the control of single quantum states, e.g., qubits, which is 
becoming important in the field of quantum computation~\cite{Monroe95}, 
we want to develop a method to control collective 
phenomena, such as phase transitions, with a laser. 

One way to control electron and atomic systems coherently is 
to exploit an interaction between matter and laser. 
Control of spin systems in condensed matter by lasers 
is becoming a realistic as well as fascinating topic 
due to recent experimental advances~\cite{Kampfrath11,Kimel05,Kirilyuk10}. 
The key in our study is to use a {\it magnetic component 
of circularly polarized lasers} 
with photon energy far below the electron energy scale. 
It was in fact demonstrated recently that a magnetic field of 
lasers in the terahertz (THz) regime can directly 
access the spin dynamics without disturbing the charge degrees of 
freedom of electrons~\cite{Kampfrath11}. 
This enables us to focus on coherent spin dynamics since 
incoherent processes arising from charge excitations can be prevented. 
While the setup of Ref.~\onlinecite{Kampfrath11} 
is within the linear-response regime, 
here we explore a nonequilibrium avenue in 
the nonperturbative regime, where we 
propose that a ``laser-induced phase transition''
with perfect quantum coherence can be realized. 
Namely, a circularly polarized laser is shown to 
induce a net magnetization in quantum antiferromagnets.
Most of the previous studies about spin pumping~\cite{Tserkovnyak02} 
or spintronics~\cite{Zutic04} 
focus on controlling some existing magnetization by a spin torque, 
which is in sharp contrast with the present nonequilibrium 
phenomenon where the magnetization rises from zero in 
a direction perpendicular to the polarization plane of the laser 
as a purely quantum-mechanical effect. 

We first demonstrate the dynamical induction of magnetization numerically with 
the infinite time-evolving block decimation 
(iTEBD)~\cite{Vidal07,Barmettler09} for 
one-dimensional spin chains. 
The iTEBD is a numerical method that exploits a matrix-product state representation, 
and can deal with infinite systems (i.e., free of finite size effects) 
by imposing a spatial periodicity. 
Through imaginary-time and real-time evolutions, 
we can obtain the GS and dynamics of a system, respectively.

Dynamical phase transitions require a theoretical 
treatment that goes beyond the linear-response theory. 
The Floquet theory is becoming a standard picture 
for studying quantum systems under 
time-periodic driving~\cite{Oka09,Kitagawa11,Lindner11,Tsuji08}.
Namely, the time periodicity enables us to cast a 
time-dependent problem into a static effective model
governed by the Floquet Hamiltonian.
This has proved to be a useful method in the theory of 
``Floquet topological insulators''~\cite{Oka09,Kitagawa11,Lindner11}, 
and has also been applied to many-body problems such as a 
photoinduced Mott transition~\cite{Tsuji08}. 
We find here that we can map spin systems in a circularly polarized laser 
onto static effective systems in a {\it slanted} magnetic field 
using unitary transformation into a rotating frame, 
or equivalently the Floquet theory. 
The emergence of the laser-induced magnetization can 
be understood from this static model, which we 
confirm by comparing exact diagonalization results with iTEBD.
Since the above discussion is also applicable to systems with 
spatial dimensions higher than one, they should accommodate 
the induction of magnetization as well.

While the magnetization can be induced in any dimensions, 
one-dimensional antiferromagnets have, quantum mechanically, 
a special interest. 
The system is known to be gapless if the size of each spin $S$ 
is a half-odd-integer, and gapped if $S$ is an integer. 
This is the celebrated Haldane's conjecture~\cite{Haldane83}, 
now established by intensive analytic, numerical, 
and experimental studies. 
Specifically, the groundstate (GS) of $S=1$ spin chain, 
known as the Haldane phase, 
is topologically protected by a symmetry~\cite{Pollmann10}, 
and is characterized by a string order parameter~\cite{Nijs89}. 
We find an interesting relationship between 
the size of the induced magnetization and 
the correlation length of the string order correlation function. 

This paper is organized as follows. 
The setup and the model is described in 
Sec.~\ref{sec:formulation}. 
We explain the mapping from the original model 
to an effective static one 
with a rotating frame in Sec.~\ref{sec:mapping}. 
Section~\ref{sec:transition} is devoted to 
an explanation about the relationship between 
the induced magnetization and 
the string order correlation function. 
A summary and discussions, including 
experimental feasibility, are given in Sec.~\ref{sec:summary}. 

\section{Formulation}
\label{sec:formulation}

\begin{figure}[t]
\includegraphics[width=0.3\textwidth]{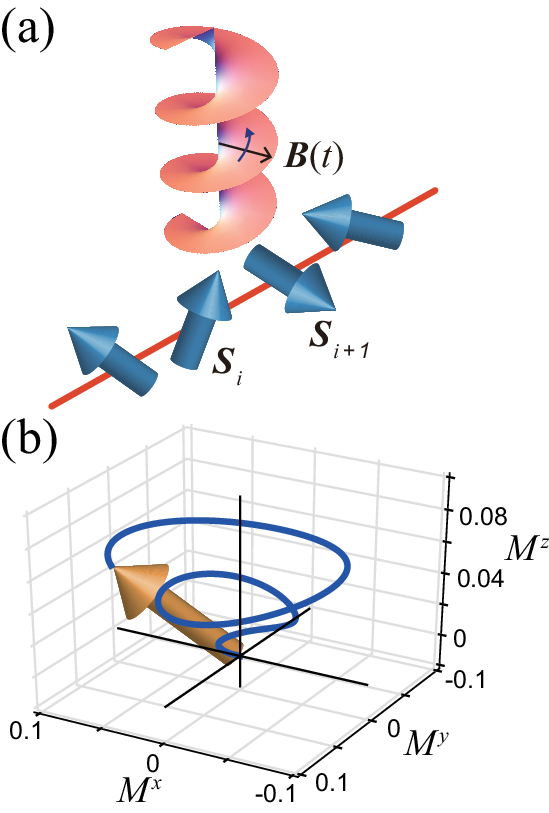}
\caption{(Color online) 
(a) A quantum spin system illuminated by a circularly polarized laser 
is schematically depicted. 
(b) A typical evolution of the magnetization induced by the laser, 
which is calculated with iTEBD for $D/J=0.25$, $A/J=1.0$, and $\Omega/J=1.4$. 
The range of the plot corresponds to about 
two cycles of the laser (a time interval $0\leq t\leq  9.6J^{-1}$). 
}
\label{fig:Setup}
\end{figure}

We consider an application of circularly polarized laser 
to quantum spin systems, as shown in Fig.~\ref{fig:Setup}(a).
In this paper, we focus on one-dimensional Heisenberg antiferromagnets 
with a Hamiltonian,
\begin{equation}
 {\cal H}_{0}=J\sum_{\langle i,j\rangle}\boldsymbol{S}_{i}\cdot\boldsymbol{S}_{j}
             +D\sum_{i}(S_{i}^{z})^{2}
             \quad(J>0),
\label{eq:Hamiltonian}
\end{equation}
where the term proportional to the coefficient $D$ is 
the magnetic anisotropy of a single-ion type. 
The model is simple, and also known to be 
realistic for describing, e.g., 
organic compounds 
such as ${\rm Ni(C_2H_8N_2)_2NO_2ClO_4}$ (NENP) and
${\rm Ni(C_5D_{14}N_2)_2N_3(PF_6)}$ (NDMAP). 
$D/J$ in NENP and NDMAP is estimated from experiments 
to be about $0.18$~\cite{Regnault94} and 0.25~\cite{Zheludev01}, respectively. 
Here, we set $D/J=0.25$. 

We consider a sudden switch-on of the laser 
at $t=0$ starting from the GS of (\ref{eq:Hamiltonian}), 
which is in the Haldane phase~\cite{Chen03}. 
The state evolves according to the Hamiltonian 
\begin{equation}
 {\cal H}(t)={\cal H}_{0} - 
A({\rm e}^{-{\rm i}\Omega t}S^{+}_{\rm tot}
                                   +{\rm e}^{ {\rm i}\Omega t}S^{-}_{\rm tot})
\label{eq:HamiltonianWithLaser}
\end{equation}
for $t>0$, where $2A$ 
and $\Omega$ are the amplitude and frequency (photon energy) 
of the laser, respectively, and
$S_{\rm tot}^{\pm}=S_{\rm tot}^{x}\pm{\rm i}S_{\rm tot}^{y}$ 
are raising and lowering operators for the total spin 
($S_{\rm tot}^{x\;(y,z)}\equiv \sum_{i}S_{i}^{x\;(y,z)}$). 
Here, we assume that only the magnetic component of the laser 
couples to the system, 
and that the laser is applied along the $z$ axis. 
Thus, each spin feels a magnetic field rotating in the $xy$ plane, 
$B(S_{\rm tot}^{x}\cos\Omega t+S_{\rm tot}^{y}\sin\Omega t)$, which 
is expressed above in terms of the spin raising and lowering 
operators ($B=2A$). 

The time evolution of the induced magnetization 
$\boldsymbol{M}$ ($\equiv\langle \boldsymbol{S}_{i}\rangle$) 
calculated with iTEBD is displayed in Fig.~\ref{fig:Setup}(b). 
Note that $\boldsymbol{M}$ is normalized so that 
fully polarized magnetization is 1. The magnetization 
starts from zero and grows in magnitude with precession. 
Remarkably, the emerging $\boldsymbol{M}$ tends to point in 
the $z$ direction, despite 
the external magnetic field being entirely in the $xy$ plane. 
There are two conditions for the emergence of 
magnetization. One is the anisotropy of the 
spin system. In SU(2) symmetric systems, 
the spin dynamics becomes trivial 
since ${\cal H}_{0}$ would then commute with 
$S^{\pm}_{\rm tot}$ and 
$S_{\rm tot}^{z}\equiv\sum_{i}S_{i}^{z}$.
In the present system, the single-ion magnetic anisotropy, 
i.e., the $D$ term, lifts this constraint.
An Ising-type anisotropy (XXZ model) has a similar effect as well. 
The second condition is to use a {\it circular} polarized light.  
In a linearly-polarized light, emergence of magnetization perpendicular to 
the applied magnetic field is prohibited by spin-inversion symmetry. 

\begin{figure}[t]
\includegraphics[width=0.3\textwidth]{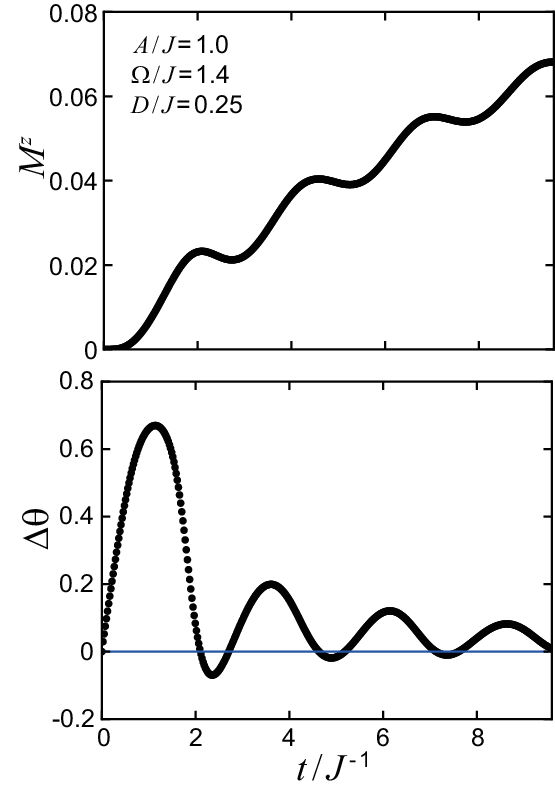}
\caption{(Color online) 
Time-evolution of magnetization along $z$-axis $M^{z}$ (top panel),
and phase difference, $\Delta \theta$ (bottom), in the $xy$ plane 
between the laser magnetic field and induced magnetization.
}
\label{fig:M_trace}
\end{figure}

As for a precession of the magnetization, 
there is an intuitive way to grasp the behavior with a 
semiclassical spin picture.  As can be seen from the numerical result 
for $M^{z}$ in Fig.~\ref{fig:M_trace}, the evolution is nonmonotonic. 
If we compare time evolution of $M^{z}$ with the relative angle 
between the directions of the external magnetic field
and magnetization: 
$\Delta\theta\equiv \Omega t-\arctan(M^{y}/M^{x})\mod\pi$, 
the regions for increasing (decreasing) $M^{z}$ are seen to 
coincide with positive (negative) $\Delta \theta$ regions. 
This indicates that $\boldsymbol{M}\equiv(M^{x},M^{y},M^{z})$ follows 
a semi-classical equation, 
$\dot{\boldsymbol{M}}=\gamma\boldsymbol{M}\times\boldsymbol{B}(t)$, 
where $\boldsymbol{B}(t)=B(\cos\Omega t,\sin\Omega t,0)$ is 
the magnetic field of the laser with $\gamma$ a positive constant. 
If we use fields with opposite circular polarization 
(left versus right), the magnetization points in the other direction. 

\begin{figure}[t]
\includegraphics[width=0.3\textwidth]{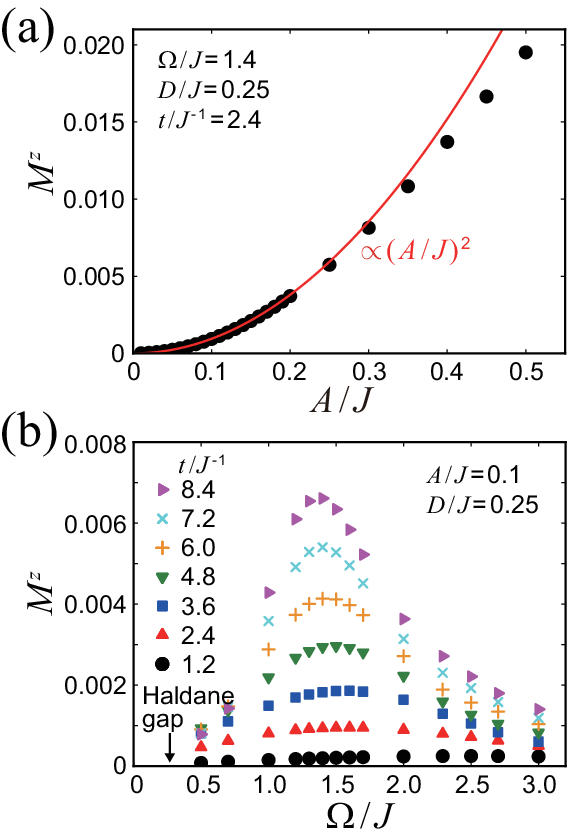}
\caption{(Color online) 
(a) $A$ dependence of $M^{z}$ for $t/J^{-1}=2.4$. 
The solid line is a fit to $M^{z}\propto (A/J)^{2}$. 
(b) $\Omega$ dependence of $M^{z}$ at various $t$. 
}
\label{fig:Mz_dep}
\end{figure}

However, the emergence of the magnetization is purely a quantum process. 
In order to clarify the mechanism, we study how
the induced magnetization depends on the laser amplitude and frequency. 
As shown in Fig.~\ref{fig:Mz_dep}(a), 
$M^{z} \propto (A/J)^{2}$ for small $A$. 
This result indicates that this is a second-order nonlinear process in terms 
of the laser magnetic field. 
Figure~\ref{fig:Mz_dep}(b) shows 
the $\Omega$ dependence of $M^{z}$ at various $t$. 
As time advances, the peak in $M^{z}$ develops 
around $\Omega/J\simeq 1.4$, which implies a resonance 
at this frequency. 
We note that this energy scale is an order of magnitude 
greater than the Haldane gap 
$\Delta/J\simeq 0.26$~\cite{Golinelli92}, 
so that this is taken to be a hallmark of the contribution 
from high-energy excited states. 
In the following, we explain the induction of magnetization and 
the resonance behavior in terms of 
mapping to an effective static model combined 
with exact diagonalization results. 

\section{Mapping to a static model}
\label{sec:mapping}

The dynamics of the present system can be understood 
by mapping it onto a system described by an effective static model 
using a unitary transformation. 
The time-dependent Schr\"odinger equation for the 
original Hamiltonian (\ref{eq:HamiltonianWithLaser}) is
\begin{equation}
\left[{\rm i}\partial_{t}-{\cal H}(t)\right]|\Psi(t)\rangle=0. 
\nonumber
\end{equation}
If we move on to a reference frame rotating with 
the magnetic component of laser using a unitary transformation 
$U=\exp({\rm i}\Omega S_{\rm tot}^{z}t)$, the new state
$|\Psi'(t)\rangle=U|\Psi(t)\rangle$ 
satisfies
\begin{equation}
U\left[{\rm i}\partial_{t}-{\cal H}(t)\right]
 U^{\dagger}|\Psi'(t)\rangle=0.\nonumber 
\end{equation}
With the commutation relations, 
\begin{align}
 \big[U,{\rm i}\partial_{t}\big]U^{\dagger}&=\Omega S_{\rm tot}^{z},
 \nonumber\\
 \big[U,{\cal H}(t)\big]U^{\dagger}&=-A\Big(
   {\rm e}^{-{\rm i}\Omega t}[U,S_{\rm tot}^{+}]
  +{\rm e}^{{\rm i}\Omega t}[U,S_{\rm tot}^{-}]
 \Big)U^{\dagger}\nonumber\\
 &=-A\left[
    S_{\rm tot}^{+}(1-{\rm e}^{-{\rm i}\Omega t})
   +S_{\rm tot}^{-}(1-{\rm e}^{ {\rm i}\Omega t})\right]\nonumber\\
 &=-2AS_{\rm tot}^{x}+A(
    {\rm e}^{-{\rm i}\Omega t}S_{\rm tot}^{+}
   +{\rm e}^{ {\rm i}\Omega t}S_{\rm tot}^{-}),\nonumber
\end{align}
the Schr\"odinger equation is cast into the form
\begin{equation}
 \left[{\rm i}\partial_{t}
   -\big({\cal H}_{0}-2AS_{\rm tot}^{x}
   -\Omega S_{\rm tot}^{z}\big)\right]|\Psi'(t)\rangle=0.
 \nonumber
\end{equation}

\begin{figure}[t]
\includegraphics[width=0.45\textwidth]{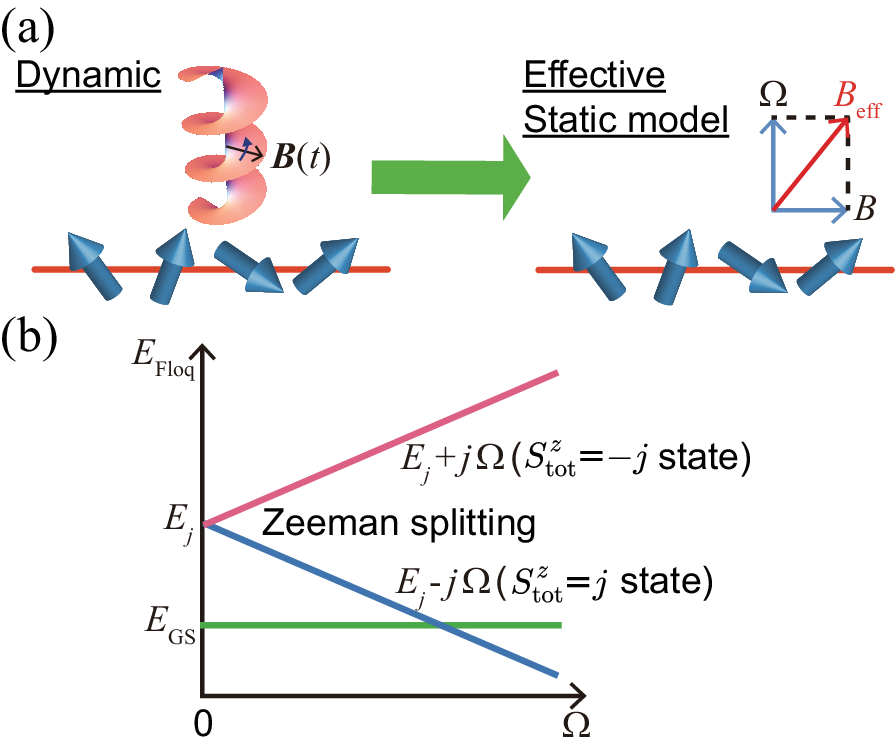}
\caption{(Color online) 
(a) The unitary transformation into a rotating frame 
maps the spin system in circularly polarized laser 
with amplitude $B$ and frequency $\Omega$
to a static spin model with a slanted magnetic field 
$\boldsymbol{B}_{\rm eff}=(B,0,\Omega)$. 
(b) An effective description of the laser-induced magnetization. 
The magnetic field in the $z$ direction $\Omega$ shifts the 
many-body energy levels with a ``Zeeman splitting.''}
\label{fig:Floquet}
\end{figure}

Thus we end up with an effective static Hamiltonian 
\begin{equation}
 {\cal H}'={\cal H}_{0}-2AS_{\rm tot}^{x}
   -\Omega S_{\rm tot}^{z},
 \label{eq:EffectiveHamiltonian}
\end{equation}
which is simply a spin system in a slanted magnetic field 
$\boldsymbol{B}_{\rm eff}=(B,0,\Omega)$ 
with $B=2A$ [Fig.~\ref{fig:Floquet}(a)].
We note that this mapping is applicable to arbitrary lattice structures 
other than the chain considered here. 
We also note that the effective description 
has resemblance with the problem of electron spin resonance 
(e.g., Ref.~\onlinecite{Oshikawa99}), where the role of the 
external magnetic field is here played by the laser frequency $\Omega$. 
Now the above treatment enables us to understand 
the mechanism of the laser-induced 
magnetization process. 
Excited states with $S_{\rm tot}^{z}=\pm j\;(j>0)$ are initially degenerate 
due to the spin inversion symmetry ($S_{i}^{z}\to-S_{i}^{z}$). 
When the laser is turned on, 
the ``longitudinal magnetic field'' $\propto \Omega$ lifts 
the degeneracy due to the Zeeman effect 
and the energy levels split into $E_{j}\to E_{j}\mp j\Omega$ 
as shown in Fig.~\ref{fig:Floquet}(b). 
The transverse field $B$ acts to hybridize the states 
having different $S_{\rm tot}^{z}$'s. 
The hybridization becomes larger as the energies come closer 
to each other. 
Since $S_{\rm tot}^{z}=+j$ states approach the GS 
while $S_{\rm tot}^{z}=-j$ states depart from it, 
the GS primarily hybridizes with $S_{\rm tot}^{z}=+j$ states, 
which is precisely why a positive net magnetization appears. 

\begin{figure}[t]
\includegraphics[width=0.3\textwidth]{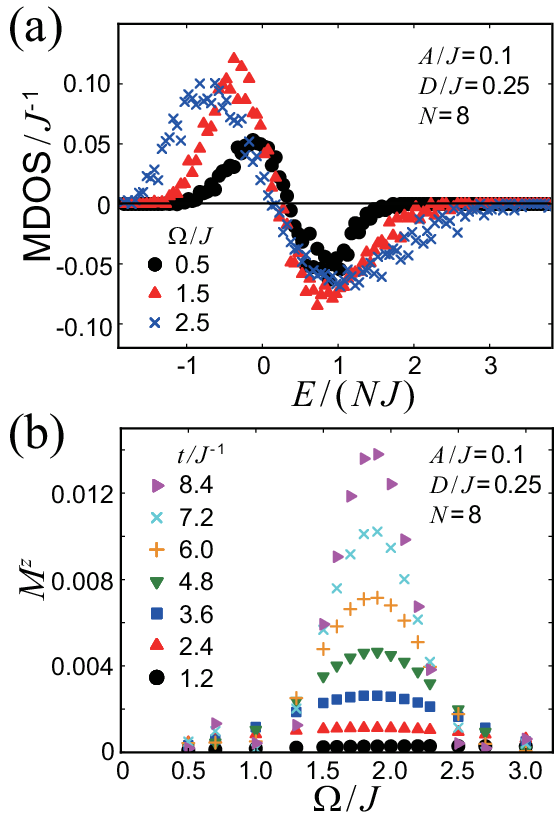}
\caption{(Color online) 
(a) Magnetic density of states (MDOS) 
in units of (number of states)$\cdot J^{-1}$ per site 
for $\Omega/J$ = 0.5, 1.5, and 2.5. 
(b) $\Omega$ dependence of $M^{z}$ at various times. 
The results are obtained by exact diagonalization of the $N=8$
effective Hamiltonian~(\ref{eq:EffectiveHamiltonian}).}
\label{fig:MDOS}
\end{figure}

\begin{figure}[t]
\includegraphics[width=0.3\textwidth]{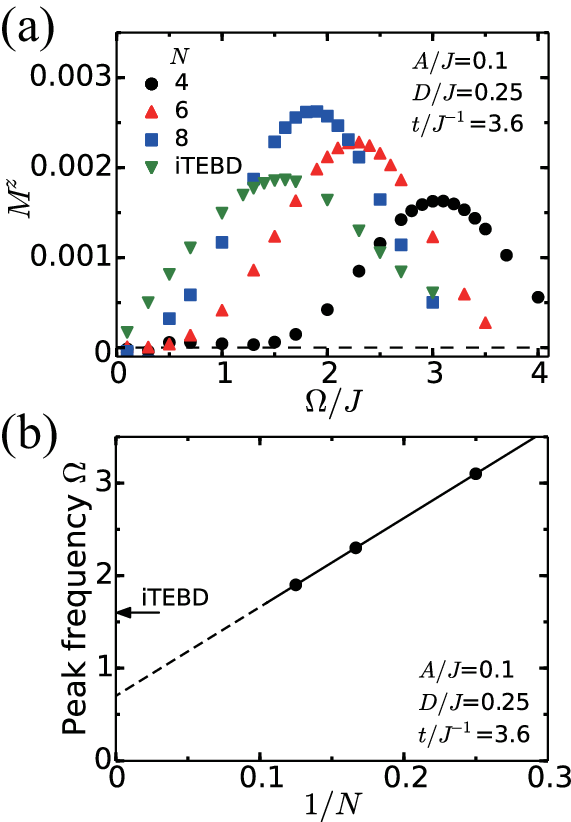}
\caption{(Color online) 
(a) $\Omega$ dependence of $M^{z}$ at $t/J^{-1}=3.6$ calculated by 
exact diagonalization for finite size systems $N=4,6,8$ 
and by iTEBD for an infinite size system.
(b) Peak frequency for $N=4,6,8$, and its linear extrapolation 
to $N\to\infty$ limit. 
The arrow represents the peak frequency obtained from iTEBD. 
}
\label{fig:FiniteSize}
\end{figure}

In the present case, the thermodynamic limit is nontrivial 
since we have to consider the hybridization 
with infinitely many levels. 
To clarify this, let us introduce the 
``magnetic density of states'' (MDOS), 
defined as $\sum_{i}M_{i}^{z}\delta(E_{i}-E)$, which 
enables us to capture the behavior of the Zeeman splitting 
for the entire many-body states. 
MDOS, normalized by $2^{N}$ ($N$ is the number of spins), 
for various values of $\Omega$ 
is calculated with diagonalization of the 
effective Hamiltonian~(\ref{eq:EffectiveHamiltonian}). 
The result for $A/J=0.1$ and $N=8$ is displayed in Fig.~\ref{fig:MDOS}(a). 
While MDOS is zero at $\Omega=0$ due to the $S_{\rm tot}^{z}=\pm j$ degeneracy, 
it develops positive (negative) peaks on the $E<0$ $(E>0)$ side, 
which move to low (high) energies with increasing $\Omega$. 
When the energy of the positive peak overlaps with
that of the GS ($E\sim -1.2NJ$), 
the hybridization becomes most prominent 
and thus leads to the resonancelike behavior. 
Figure ~\ref{fig:MDOS}(b) plots 
$M^{z}$ against $\Omega$ at various times 
calculated by combining the Floquet theory and 
the exact diagonalization for $N=8$ (Appendix~\ref{sec:Floquet}). 
We can see that the result agrees well 
with the iTEBD result [Fig.~\ref{fig:Mz_dep}(b)]. 
The peak is located around $\Omega/J\simeq 1.9$, which is close to 
the peak around $\Omega/J\simeq 1.4$ in the iTEBD result. 
A slight deviation can be attributed to a finite-size effect. 
We can compare 
the $\Omega$ dependence of $M^{z}$ at $t/J^{-1}=3.6$ calculated by 
exact diagonalization for finite size systems $N=4,6,8$ 
with the iTEBD result 
for an infinite system in Fig.~\ref{fig:FiniteSize}(a). 
As we increase $N$, the exact diagonalization result 
approaches that of iTEBD qualitatively. 
However, a deviation is seen when we use a naive linear extrapolation 
in the limit of $N\to\infty$ [Fig.~\ref{fig:FiniteSize}(b)]. 

\begin{figure}[t]
\includegraphics[width=0.3\textwidth]{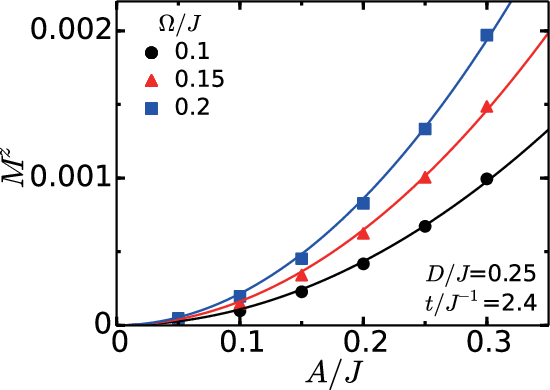}
\caption{(Color online) 
The induced $M^{z}$ for small $A$ and $\Omega$. 
Lines represent fitting of the data points 
by parabolic curves.
}
\label{fig:Gap}
\end{figure}

Finally in this section, 
we note that the emerged magnetization represents 
not a quantum phase transition 
but a dynamical phase transition. 
Due to the sudden onset of the laser (an ``ac quench''), 
the state becomes highly excited. 
Since the present unitary time evolution described by the 
static effective model keeps the energy unchanged, 
the system after the quench does not decay to the GS.
Let us display the induced $M^{z}$ for small $A$ and $\Omega$ in 
Fig.~\ref{fig:Gap}. 
We can see that the induced $M^{z}$ grows as $(A/J)^{2}$, 
while in the effective model~(\ref{eq:EffectiveHamiltonian}) 
the transition does not occur when the size of 
the effective field $\sqrt{(2A)^{2}+\Omega^{2}}/J$ is smaller than 
the Haldane gap $\Delta/J\sim 0.26$. 
The result implies that the magnetization induction is 
a dynamical phase transition and not a static quantum phase transition 
at the absolute zero $T$. 

\section{Breakdown of the string order}
\label{sec:transition}
\begin{figure}[t]
\includegraphics[width=0.3\textwidth]{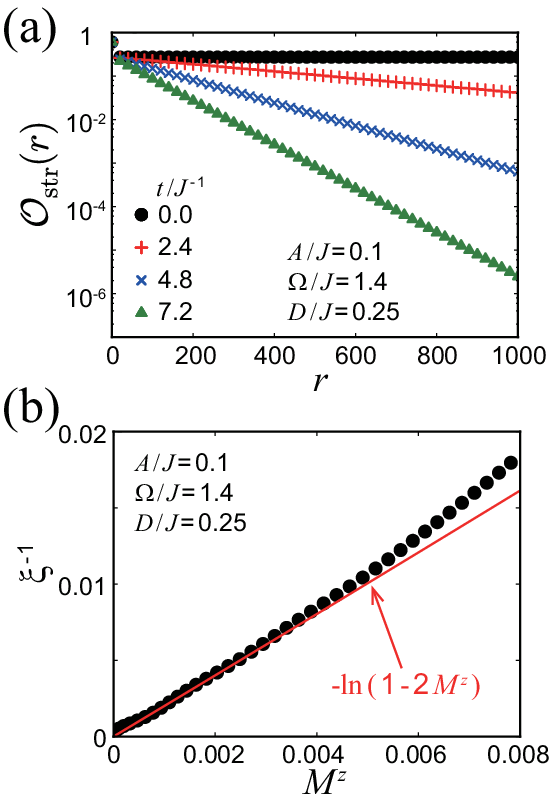}
\caption{(Color online) 
(a) The string order correlation function 
calculated with iTEBD at various times. 
(b) The inverse of string correlation length $\xi^{-1}$ plotted 
as a function of $M^{z}$. The result of iTEBD (circles) and 
the analytical prediction, $\xi^{-1}=-\ln(1-2M^{z})$ (line) are shown.}
\label{fig:Haldane}
\end{figure}

In this section, we present an interesting relation 
between the induced magnetization and 
the string order correlation~\cite{Nijs89}. 
The Haldane phase, being topologically ordered, has no 
local order parameter, 
but is instead characterized by the string order parameter, 
which is defined as $\lim_{r\to\infty}{\cal O}_{\rm str}(r)$ 
from the string-order correlation function, 
${\cal O}_{\rm str}(r)=
\langle S_{0}^{z}\exp({\rm i}\pi\sum_{i=0}^{r-1}S_{i}^{z})
        S_{r}^{z}\rangle$~\cite{Nijs89}. 
The string order parameter can capture the Haldane phase 
since this phase consists of a superposition of states 
in which an arbitrary number of $|S_{i}^{z}=0\rangle$'s are 
inserted between $|1\rangle$ and $|-1\rangle$ in a N\'eel order 
in terms of the $S_{i}^{z}$ basis, 
e.g., $|\ldots,1,0,0,0,-1,0,1,0,0,-1,\ldots\rangle$.
In the initial GS, we are in the Haldane phase with the 
nonzero string order parameter as seen 
in the $t=0$ result of Fig.~\ref{fig:Haldane}(a). 
For $t>0$, the string order correlation 
${\cal O}_{\rm str}(r)$ decays exponentially with distance $r$.
This result indicates that the breakdown of the string order happens 
as soon as the laser application begins. 

We find that the decay of the 
string order has a direct (and even analytic) relation with the 
emergence of the $z$ component of magnetization. 
Specifically, the magnetization is related 
with the string correlation length $\xi$ 
[defined by fitting ${\cal O}_{\rm str}(r) \propto \exp(-r/\xi)$]. 
Figure ~\ref{fig:Haldane}(b) plots $\xi^{-1}$ as a function of $M^{z}$. 
We can deduce an analytic form for the relation 
between $\xi$ and $M^{z}$ as follows. 
A flip of a single spin from $|0\rangle$ to $|1\rangle$ 
or from $|-1\rangle$ to $|0\rangle$ in the Haldane phase 
gives a factor $-1$ in the string order correlation function, and
acts like a ``disorder'' in the string-ordered states
(e.g., 
$|\ldots,1,0,0,0,-1,0,1,0,0,-1,\ldots\rangle\to
|\ldots,1,0,$``1''$,0,-1,0,1,0,0,-1,\ldots\rangle$). 
The probability of having such a disorder is $M^{z}$ per site, 
and by estimating the probability of having $k$ disordered sites out of $r$
sites, we are led to an expression, 
\begin{align}
{\cal O}_{\rm str}(r)
  &\simeq S_{0}\sum_{k=0}^{r}(-1)^{k}{\,}_{r}{\rm C}_{k}(M^{z})^{k}(1-M^{z})^{r-k} 
  \nonumber\\
  &=      S_{0}(1-2M^{z})^{r},
\nonumber
\end{align}
where $S_{0}$ is the initial (i.e., at $M^{z}=0$) 
value of the string order.  Namely, we have $\xi^{-1}=-\ln(1-2M^{z})$, 
which agrees well with numerical data 
in Fig.~\ref{fig:Haldane}(b) in the small $M^{z}$ region. 

Now we can comment on the symmetry protection and 
string order of the Haldane phase. 
In Ref.~\onlinecite{Pollmann10},
it is proved that the Haldane phase is characterized by 
a twofold degeneracy in the entanglement spectrum, and 
that this degeneracy can be protected by imposing any one of 
(i) time-reversal, (ii) bond-center inversion, 
or (iii) $Z_{2}\times Z_{2}$ symmetries.
It is also discussed that 
the string order parameter is well defined in systems 
with $Z_{2}\times Z_{2}$ symmetry. 
An example of Haldane phases with no string order is given 
in Ref.~\onlinecite{Gu09}. 
In our case, however, the perturbation is single-ion anisotropy 
and magnetic field. 
The single-ion anisotropy, which is unchanged under $S_{i}^{z}\to -S_{i}^{z}$, 
does not break $Z_{2}\times Z_{2}$ symmetry. 
This is consistent with our data for $t=0$ 
in Fig.~\ref{fig:Haldane} (a), which shows 
a nonzero string order parameter. 
Although the string order parameter cannot be used as an order parameter 
in the magnetized phase, 
our finding shows that the decay of the string order correlation function 
is related with the size of the induced magnetization. 

\section{Summary and discussions}
\label{sec:summary}

To summarize, we have found that 
a net magnetization can be induced by applying 
a circularly polarized laser to antiferromagnets with anisotropy. 
We numerically demonstrated this phenomenon using iTEBD. 
The mechanism of the phenomenon and the existence of 
a resonant frequency are explained 
by a unitary transformation into a rotating frame, 
equivalently by the Floquet theory. 
Then the system is described by an effective static picture, 
in which the amplitude and frequency of the laser act as 
transverse and longitudinal magnetic fields, respectively. 
While the $S=1$ spin chain is initially in the Haldane phase, 
the laser-induced magnetization occurs concurrently 
with a destruction of the topological order, 
which is characterized by an exponential decay 
of the string order correlation function. 

Let us discuss the experimental feasibility. 
Pump-probe experiments should be most 
promising, where the pump laser is required to be 
strong and circularly polarized, 
and the induced magnetization can be 
measured by Faraday or Kerr rotation. 
The necessary pump strength depends on the 
sensitivity of the probe. 
We make an estimation for NDMAP,  
whose exchange interaction is estimated to be 
$J \simeq 2.8$ meV with $D/J\sim 0.25$~\cite{Zheludev01}. 
While we have considered a continuous laser application, 
a pulse containing only a few cycles of laser can also 
induce a magnetization as illustrated in Fig.~\ref{fig:Setup}(b). 
In order to induce $M_{z}\sim 0.01$ with few cycles, 
we need $A/J\sim 0.1$. 
This corresponds to $B=2A=0.2J\simeq 5.6$ T. 
The optimum frequency is, from 
the resonance frequency, estimated to be $1.4J\simeq 3.9$ meV$\sim 1$ THz. 
The currently available intensity of the magnetic field in a laser 
is $\sim 0.13$ T~\cite{Kampfrath11}, which is still an order 
of magnitude smaller than the required strength. 
However, THz laser techniques are rapidly 
advancing~\cite{Matsunaga13,Hirori11,Kirilyuk10,Vicario13}, 
and we expect our theoretical proposal to become experimentally 
feasible in the near future. 

Finally, we comment on the scope of the laser-induced magnetization. 
While we have presented the result for one-dimensional spin-1 systems, 
our discussion does not depend on the dimension nor the size of the spin; 
spin-related phenomena abound in higher dimensions, 
such as the effect of frustration 
and emergence of spin liquid phase to name a few. 
Moreover, quantum spin systems have interdisciplinary spin-offs, 
e.g., cold-atom systems. 
It is an interesting future problem to study 
laser-induced phase transitions in such systems, 
and the theory presented here is expected to play 
an important role. 

\acknowledgements
We acknowledge illuminating discussions with 
S. C. Furuya, N. Tsuji, and P. Werner. 
The computation was partially performed on computers at the Supercomputer
Center, Institute for Solid State Physics, the University of Tokyo.
This work is supported in part by Grants-in-Aid from
JSPS, Grant No. 23740260.

\appendix
\section{Floquet theory}
\label{sec:Floquet}

Here let us show that the 
effective static model~(\ref{eq:EffectiveHamiltonian}) 
can equivalently be obtained using the Floquet theory for many-body systems 
instead of the unitary transformation onto the rotating frame.
The Floquet theory is a mathematical technique to treat time-periodic 
differential equations, which is a temporal analog of 
the Bloch theorem for spatially periodic systems.
When applied to a time-dependent Schr\"odinger equation,
\begin{equation}
{\rm i}\partial_{t}|\Psi(t)\rangle={\cal H}(t)|\Psi(t)\rangle, 
\label{eq:TdepSchrodinger}
\end{equation}
with time periodicity ${\cal H}(t+T)={\cal H}(t)$ ($T$: period), 
the Floquet theorem dictates that the solution 
should have a form $|\Psi(t)\rangle={\rm e}^{-{\rm i}\epsilon t}|\Phi(t)\rangle$, which is a product of 
a phase factor involving $\epsilon$ called Floquet quasi-energy and 
a time-periodic wave function (Floquet state) with $|\Phi(t+T)\rangle=|\Phi(t)\rangle$.   
With both ${\cal H}(t)$ and $|\Phi(t)\rangle$ periodic in $t$, 
we can make a discrete Fourier transform,
\begin{align}
{\cal H}(t)&=\sum_{m}{\rm e}^{-{\rm i}m\Omega t}H_{m},\nonumber\\ 
|\Phi(t)\rangle&=\sum_{m}{\rm e}^{-{\rm i}m\Omega t}|\Phi^{m}\rangle.\nonumber 
\end{align}
When these are plugged into Eq.~(\ref{eq:TdepSchrodinger}), 
the Schr\"odinger equation is casted into 
a time-independent eigenvalue equation in a matrix form, 
\begin{equation}
 \sum_{m}(H_{n-m}-m\Omega\delta_{mn})|\Phi^{m}\rangle
                            =\epsilon|\Phi^{n}\rangle,
\label{eq:FloquetEigen}
\end{equation}
which can be thought of as an equation for 
``photon-dressed'' (Floquet) modes.

\begin{figure}[t]
\includegraphics[width=0.3\textwidth]{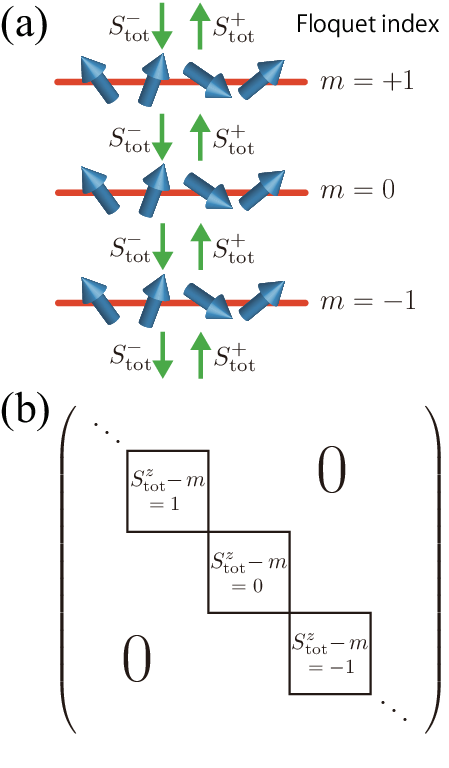}
\caption{(Color online) 
(a) Structure of the Floquet Hamiltonian (\ref{eq:FloquetEigenMatrix})
for spin systems in a circularly polarized light field. 
(b) Block-diagonal structure of the 
Floquet Hamiltonian reshuffled in terms of the good quantum
number $S_{\rm tot}^{z}-m$.}
\label{fig:FloquetConcept}
\end{figure}

The present Hamiltonian is a one-dimensional 
Heisenberg antiferromagnet with a single-ion anisotropy 
in a circularly polarized field: 
\begin{equation}
 {\cal H}(t)=J\sum_{\langle i,j\rangle}\boldsymbol{S}_{i}\cdot\boldsymbol{S}_{j}
            +D\sum_{i}(S_{i}^{z})^{2}
            -A({\rm e}^{-{\rm i}\Omega t}S_{\rm tot}^{+}
                      +{\rm e}^{ {\rm i}\Omega t}S_{\rm tot}^{-}).
\label{eq:OrigModel}
\end{equation}
Thus the zeroth component ($H_{n-m}$ with $n=m$) is 
$H_{0}=J\sum_{\langle i,j\rangle}\boldsymbol{S}_{i}\cdot\boldsymbol{S}_{j}+D\sum_{i}(S_{i}^{z})^{2}$, 
and $H_{\pm 1}$ just corresponds to $-AS_{\rm tot}^{\pm}$. 
The eigenvalue equation~(\ref{eq:FloquetEigen}) then 
simplifies into a tridiagonal form,
\begin{widetext}
\begin{equation}
\begin{pmatrix}
\;\ddots&&&&&&\\
&H_{0}-2\Omega&H_{+1}&0&0&0&\\
&H_{-1}&H_{0}-\Omega&H_{+1}&0&0&\\
&0&H_{-1}&H_{0}&H_{+1}&0&\\
&0&0&H_{-1}&H_{0}+\Omega&H_{+1}&\\
&0&0&0&H_{-1}&H_{0}+2\Omega&\\
&&&&&&\ddots\;
\end{pmatrix}
\begin{pmatrix}
\vdots\\
|\Phi^{ 2}\rangle\\
|\Phi^{ 1}\rangle\\
|\Phi^{ 0}\rangle\\
|\Phi^{-1}\rangle\\
|\Phi^{-2}\rangle\\
\vdots
\end{pmatrix}
=\epsilon
\begin{pmatrix}
\vdots\\
|\Phi^{ 2}\rangle\\
|\Phi^{ 1}\rangle\\
|\Phi^{ 0}\rangle\\
|\Phi^{-1}\rangle\\
|\Phi^{-2}\rangle\\
\vdots
\end{pmatrix},
\label{eq:FloquetEigenMatrix}
\end{equation}
\end{widetext}
where $\Omega$ is a shorthand for $\Omega$ times the unit matrix. 
The Floquet formalism can then be schematically depicted in 
Fig.~\ref{fig:FloquetConcept}(a), where 
replicas of the original system are prepared 
for different Floquet (photon-dressed) modes $m$. 
The terms with the phase factor 
${\rm e}^{\mp{\rm i}\Omega t}$ in the Hamiltonian 
induce a transition from the Floquet mode $m$ to $m\pm 1$. 

The matrix representation of Eq.~(\ref{eq:FloquetEigen}) 
is in general infinite-dimensional, since $m$ takes 
from $-\infty$ to $+\infty$. 
In the present case, however, 
$S_{\rm tot}^{z}-m$ 
is a good quantum number because the term 
$S_{\rm tot}^{\pm}$ appears in the Hamiltonian 
with a phase factor ${\rm e}^{\mp{\rm i}\Omega t}$, so that 
the term simultaneously changes the Floquet index and $S_{\rm tot}^{z}$ 
by $\pm 1$, respectively, 
as shown in Fig.~\ref{fig:FloquetConcept}(a). 
This implies that the Floquet matrix can be put into a 
block-diagonal form as shown in Fig.~\ref{fig:FloquetConcept}(b). 
We call the Hamiltonian that acts within the 
blocks an ``irreducible Floquet Hamiltonian.''
If the system size (i.e., the total number of spins $N$) is finite, 
each block is finite-dimensional 
since $S_{\rm tot}^{z}$ is bounded as $-N\leq S_{\rm tot}^{z}\leq N$. 
Thus, even in the presence of the time-dependent 
external field, we can readily solve the eigenvalue equation 
with an exact diagonalization as far as finite systems 
are concerned. 
The necessary and sufficient conditions 
to obtain an irreducible Floquet Hamiltonian are
(i) $[H_{0},S_{\rm tot}^{z}]=0$ and 
(ii) conservation of $S_{\rm tot}^{z}-m$ ($m$: Fourier mode index). 
These conditions imply that the direction of laser propagation 
should be parallel to the anisotropy axis of the magnet. 
This dictates the experimental setup. 

The irreducible Floquet Hamiltonian is formally equivalent to the 
Hamiltonian for an $S=1$ chain with longitudinal 
and transverse magnetic fields 
in the present spin model~(\ref{eq:OrigModel}). 
Namely, since the $S_{\rm tot}^{\pm}$ term connects the sectors 
that have $S_{\rm tot}^{z}$ differing by $\pm 1$, 
$B(=2A)$ acts as the transverse magnetic field. 
On the other hand, $\Omega$ acts as the longitudinal magnetic field. 
We can see this 
because the matrix components in the same $S_{\rm tot}^{z}$ sector are 
$H_{0}-m\Omega$, which translates into $H_{0}-\Omega S_{\rm tot}^{z}$ 
since $S_{\rm tot}^{z}-m$ is constant 
within each irreducible Floquet Hamiltonian 
(where the constant, being irrelevant, can be set to 0). 
We end up with an effective Hamiltonian,
\begin{equation}
H_{\rm Ir. Fl.}=J\sum_{\langle i,j\rangle}\boldsymbol{S}_{i}\cdot\boldsymbol{S}_{j}
               +D\sum_{i}(S_{i}^{z})^{2}
               -BS_{\rm tot}^{x}-\Omega S_{\rm tot}^{z}.
\label{eq:FloquetHamiltonian}
\end{equation}

The time evolution of $M^{z}$ can then be calculated 
from the exact diagonalization result for 
the eigenvalues $\{\epsilon_{\alpha}\}$ 
and eigenvectors $\{|\Phi_{\alpha}\rangle\}$ 
of Eq.~(\ref{eq:FloquetHamiltonian}).  
We can reconstruct the solution of the original time-dependent 
Schr\"odinger equation as 
$|\Psi(t)\rangle = \sum_{\alpha}c_{\alpha}|\Psi_{\alpha}(t)\rangle$, 
where $|\Psi_{\alpha}(t)\rangle\equiv\sum_{m}
{\rm e}^{-{\rm i}(\epsilon_{\alpha}+m\Omega)t}|\Phi_{\alpha}^{m}\rangle$, 
and the coefficients $c_{\alpha}$, with 
$\sum_{\alpha}|c_{\alpha}|^{2}=1$ for normalization, 
is determined from the initial condition. 
For example, when the application of laser begins suddenly at $t=0$, 
the coefficients are
$c_{\alpha}=\langle\Psi_{\alpha}(t=0)|\Psi_{0}\rangle$, 
where $|\Psi_{0}\rangle$ is the initial state, 
i.e., the GS of 
${\cal H}_{0}=J\sum_{\langle i,j\rangle}\boldsymbol{S}_{i}\cdot\boldsymbol{S}_{j}+D\sum_{i}(S_{i}^{z})^{2}$. 
The magnetization per spin $M^{z}$ then evolves as 
\begin{align}
 M^{z}(t)&=\sum_{\alpha,m}|c_{\alpha}|^{2}
           \langle\Phi_{\alpha}^{m}|S_{\rm tot}^{z}/N|\Phi_{\alpha}^{m}
           \rangle\nonumber\\
         &+\sum_{\alpha<\beta,m}[c_{\beta}^{*}c_{\alpha}
           {\rm e}^{{\rm i}(\epsilon_{\beta}-\epsilon_{\alpha})t}
           \langle\Phi_{\beta}^{m} |S_{\rm tot}^{z}/N|\Phi_{\alpha}^{m}\rangle
          +{\rm H.c.}],
\nonumber
\end{align}
where the first (second) term is the $t$-independent ($t$-dependent) part. 

\section{An animation of the magnetization emergence}

\begin{figure}[t]
\includegraphics[width=0.3\textwidth]{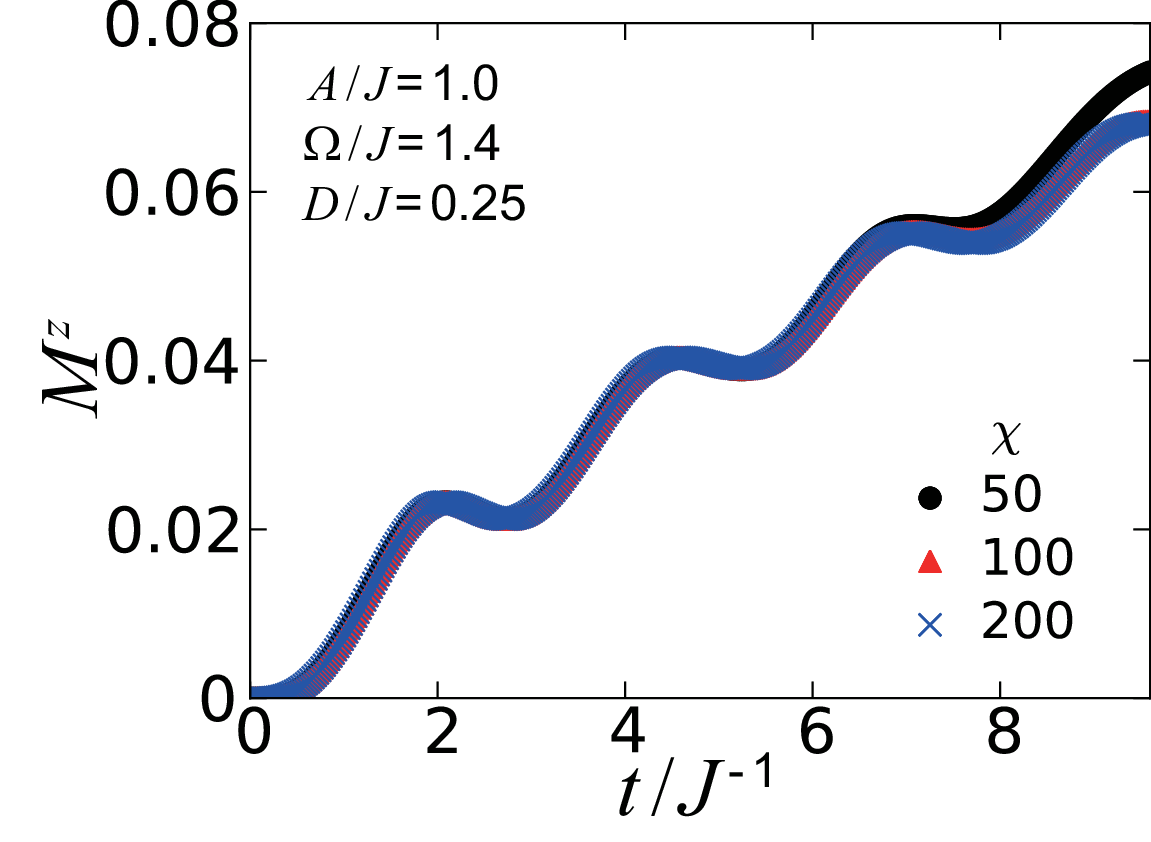}
\caption{(Color online) 
The time evolution of $M^{z}$ calculated by iTEBD 
for the matrix dimension varied as $\chi=50$, 100, 200.}
\label{fig:chidep}
\end{figure}

As a best way to represent the time-evolution of the 
magnetization, we attach here a movie~\cite{Supplement}. 
There, the initial state ($t=0$) is the GS of 
\begin{equation}
{\cal H}_{0}=J\sum_{\langle i,j\rangle}\boldsymbol{S}_{i}\cdot\boldsymbol{S}_{j}
            +D\sum_{i}(S_{i}^{z})^{2}
            \quad(J>0),\nonumber
\end{equation}
and the time evolution for $t>0$ is obtained with 
iTEBD~\cite{Vidal07} for the Hamiltonian, 
\begin{equation}
 {\cal H}(t)={\cal H}_{0}
            -A({\rm e}^{-{\rm i}\Omega t}S_{\rm tot}^{+}
              +{\rm e}^{ {\rm i}\Omega t}S_{\rm tot}^{-}),\nonumber
\end{equation}
where we have set $D/J=0.25$, $A/J=1.0$, and $\Omega/J=1.4$. 
In the movie, the time evolution of magnetization 
$\boldsymbol{M}=(M^{x},M^{y},M^{z})$ 
($M^{\alpha}\equiv\langle S_{i}^{\alpha}\rangle$) 
is represented by an arrow for a time interval of $0\leq t\leq 9.6J^{-1}$.

Let us mention the precision of iTEBD calculations, 
which is determined by the dimension $\chi$ 
of the matrix product state representation. 
The $\chi$ dependence is shown in Fig.~\ref{fig:chidep}. 
As seen, the $\chi$ dependence is small for $\chi\geq 100$. 
Thus the calculations for the time interval considered above
should be accurate since we have set $\chi=200$ 
for all iTEBD calculations in our paper.

\end{document}